\documentclass[11pt]{article}

\makeatletter\renewcommand{\section}{\@startsection
{section}{1}{\z@}{-3.5ex plus -1ex minus
    -.2ex}{2.3ex plus .2ex}{\bf }}

\makeatletter\renewcommand{\subsection}{\@startsection{subsection}{2}{\z@}{-3.25ex
plus -1ex minus
   -.2ex}{1.5ex plus .2ex}{\it }}
\makeatletter\renewcommand{\subsubsection}{\@startsection{subsubsection}{3}{-2.45ex}{-3.25ex
plus -1ex minus -.2ex}{1.5ex plus .2ex}{\it }}

\makeatletter \@addtoreset{equation}{section}

\usepackage{graphicx}
\usepackage{epsfig}

\newcommand{\be}{\begin{equation}}
\newcommand{\ee}{\end{equation}}
\newcommand{\bea}{\begin{array}}
\newcommand{\ea}{\end{array}}
\newcommand{\beqa}{\begin{eqnarray}}
\newcommand{\eeqa}{\end{eqnarray}}
\newcommand{\nn}{\nonumber}

\renewenvironment{thebibliography}[1]
     {\baselineskip=16pt plus 2pt minus 1pt
      \section*{\large\refname
        \@mkboth{\MakeUppercase\refname}{\MakeUppercase\refname}}%
     \list{\@biblabel{\@arabic\c@enumiv}}%
           {\settowidth\labelwidth{\@biblabel{#1}}%
            \leftmargin\labelwidth
            \advance\leftmargin\labelsep
            \@openbib@code
            \usecounter{enumiv}%
            \let\p@enumiv\@empty
            \renewcommand\theenumiv{\@arabic\c@enumiv}}%
      \sloppy
      \clubpenalty4000
      \@clubpenalty \clubpenalty
      \widowpenalty4000%
      \sfcode`\.\@m}

\let\fn\footnote
\renewcommand{\footnote}[1]{\linespread{1.1}\fn{#1}\linespread{1.29}}

\usepackage{amsfonts}
\usepackage{amssymb}
\usepackage{mathrsfs}
\usepackage{amsmath,amssymb}

\def\tyng(#1){\hbox{\tiny$\yng(#1)$}}

\begin{document}

\begin{titlepage}
\begin{flushright}
ITP-UH-16/08\\
\end{flushright}
\vskip 2.0cm

\begin{center}

\centerline{{\Large \bf Noncommutative $Q$-Lumps}}

\vskip 2em

\centerline{\large \bf Se\c{c}kin~K\"{u}rk\c{c}\"{u}o\v{g}lu}

\vskip 2em

{\small \centerline{\sl Institut f\"ur Theoretische Physik,
Leibniz Universit\"at Hannover} \centerline{\sl Appelstra\ss{}e 2,
D-30167 Hannover, Germany}

\vskip 1em

{\sl  e-mail:}  \hskip 2mm {\sl
seckin.kurkcuoglu@itp.uni-hannover.de} }
\end{center}

\vskip 2cm

\begin{quote}
\begin{center}
{\bf Abstract}
\end{center}
\vskip 2em
$Q$-lumps associated with the noncommutative ${\mathbb C}P^N$ model in $2+1$ dimensions are constructed. These are solitonic configurations which are time dependent and rotate with constant angular frequency. Energy of the $Q$-lumps is
$E=2 \pi k + \alpha |Q|$, and  we find that in a regime in which the noncommutativity parameter $\theta$ is related to the moduli 
determining the size of the lumps, it can be viewed to depend on $\theta$ via the Noether charge $Q$. We present a collective coordinate-type analysis signalling that ${\mathbb C}P^1$ $Q$-lumps remain stable under small radiative perturbations.
\vskip 5pt

Pacs: 11.10.Lm, 11.10.Nx, 02.40.Gh
\end{quote}

\vskip 1cm
\begin{quote}
\end{quote}

\end{titlepage}

\setcounter{footnote}{0}

\newpage

\section{{\bf Introduction}}

It was shown by Leese \cite{Leese} quite some time ago that the
${\mathbb C}P^1$ model in $2+1$ dimensions modified by the addition of a certain potential term
admits $Q$-ball \cite{Coleman} type solutions. These are time dependent 
configurations with conserved topological charge (winding number)
and Noether charge, and named $Q$-lumps in the literature
\cite{Leese}. $Q$-lumps of the ${\mathbb C}P^1$ model are constructed by
finding a simple extension of the BPS equations. Existence of this 
BPS-type bound ensures finite energy solutions, which are determined in
terms of the Noether charge $Q$, the coupling constant of the
potential and the winding number $k \in {\mathbb Z}$. For finiteness of the 
energy, it is necessary that $k \geq 2$, since the Noether charge of the  
$k=1$ configurations diverges logarithmically. However, $k=1$ lumps can exist 
as a part of a configuration of multilumps. $Q$-lump configurations in the
${\mathbb C}P^N$ models have also been studied in the literature \cite{Abraham}.
They appear as stationary solutions of K\"{a}hler sigma models modified by a
potential term which is left invariant under the transformations
generated by a Killing vector of the target manifold.
In general, the moduli space of the ${\mathbb C}P^N$ $Q$-lumps is
smaller than that of the pure ${\mathbb C}P^N$ model lumps,
since a solution with a given value of $Q$ may be scaled to give
another solution with a different value of $Q$ \cite{Leese, Abraham}.
It was also found that $(4,4)$-supersymmetric $1+1$-dimensional sigma
models with hyper-K\"{a}hler target spaces admit $Q$-kink solutions. 
These are stationary configurations, which also saturate a BPS-type bound
and carry $1/2$ of the supersymmetry \cite{Abraham-Townsend}. More recent
investigations indicated that ${\cal N}=2$ supersymmetric four-dimensional
hyper-K\"{a}hler sigma models have $Q$-lump configurations which are $1/4$ or $1/8$
BPS states \cite{Gauntlett-Tong-Townsend, Naganuma-Nitta-Sakai}. $Q$-lumps of the 
${\cal N}=2$ supersymmetric ${\mathbb C}P^N$ model carrying half of the supersymmetries
has appeared in \cite{Bak}.

Noncommutative(NC) field theories have been under investigation for
about a decade now. Among them, field theories defined on the Groenewold-Moyal(GM) type deformations
of spacetime (i.e. the noncommutative algebra ${\mathcal A}_\theta({\mathbb R}^{(d+1)})$) 
hold a considerably large part of the literature. (See, for instance \cite{Nekrasov, Szabo} for comprehensive reviews).  
Formulation of instantons and solitons on the GM spacetime and other
noncommutative spaces, such as the noncommutative tori and fuzzy
spaces, have been extensively studied and found to present very
rich mathematical structures \cite{Nekrasov, Szabo, Harvey1, Fuzzy}. 
${\mathbb C}P^N$ models on the GM spacetime have been formulated and their
BPS configurations were found in \cite{Lee}. 
Stability properties of these models as well as the $U(N)$ chiral model on the GM spacetime
have been studied in considerable detail in \cite{Lechtenfeld-Domrin}. NC $Q$-balls were investigated
in \cite{QBalls}. 

It is therefore desirable to explore the $Q$-lump configurations associated with the 
NC ${\mathbb C}P^N$ models. It appears to be rather straightforward to construct these configurations
and it turns out that, they strongly resemble their commutative cousins, and they too rotate with constant 
angular frequency in time. Nevertheless, we also find that in a regime in which the noncommutativity parameter 
$\theta$ is related to the moduli determining the size of the lumps, the energy of the NC $Q$-lump configurations 
can be viewed to depend on $\theta$ via the Noether charge $Q$.

In the next section, $Q$-lump configurations of the NC ${\mathbb C}P^N$ model is presented.
In Sec. 3, we focus on the ${\mathbb C}P^1$ model $Q$-lumps and first see that at the 
elementary classical level their stability properties are similar to that of the commutative model. 
Subsequently, we present a collective-coordinate-type analysis signalling that NC ${\mathbb C}P^1$ model $Q$-lumps
remain stable under small radiative perturbations. Contrary to the behavior of commutative $Q$-lumps, we
find that the period of fluctuations around the radially symmetric
$Q$-lump configurations depend, in addition to $\alpha$, on a function
$A(\kappa_0)$ of the ratio of the initial size of the lump to the
scale of the noncommutativity parameter $\kappa_0 := \frac{\lambda_0}{\sqrt
{2 \theta}}$. We discuss this and other related findings in some detail and compare it with the properties 
of the commutative theory, and show that our results go smoothly to those of the latter as $\theta$ tends to zero.

Throughout this paper, we work on the Groenewald-Moyal spacetime ${\mathcal A}_\theta({\mathbb R}^{2+1})$
defined by the commutation relations
\be
\lbrack {\hat x}_\mu \,, {\hat x}_\nu \rbrack = i \theta_{\mu \nu} \,, \quad \mu \,, \nu = 0,1,2 \,,
\ee
and assume that the spatial coordinates commute with time $t=x_0$, i.e. $\theta_{0i} = 0$.

\section{{\bf Q-lumps of the NC ${\mathbb C}P^N$ Model}}

To facilitate the construction of the NC ${\mathbb C}P^N$ $Q$-lumps, we start with the Lagrangian
\be
L = 2 \pi \theta \frac{1}{2} Tr \partial_\mu P \partial^\mu P + V(P) \,,
\label{eq:Lagrangian}
\ee
where $P$ is a projector living in the space ${\mathcal A}_\theta({\mathbb R}^{2+1}) \otimes Mat(N+1)$ and we
also have that $Tr = Tr_{{\cal F}} \otimes Tr_{Mat(N+1)}$, where ${\cal F}$ is the standard Fock space. 

The ${\mathbb C}P^N$ manifold is defined through the $(N+1)$-component complex unit vector 
\be
\chi = \left (
\begin{array}{c}
u  \\
1
\end{array}
\right ) \frac{1}{\sqrt{u^\dagger u + 1 }} \,,
\quad u \equiv 
\left (
\begin{array}{c}
u_1 \\
 u_2 \\
\vdots \\
 u_N
\end{array}
\right )
\,, \quad \chi^\dagger \chi = 1 \,.
\ee
$\chi$ is the partial isometry associated with the projector $P$ via $P = \chi \chi^\dagger$.

The potential term $V(P)$ in (\ref{eq:Lagrangian}) may be given as 
\be
V(P) = - 2 \pi \theta \frac{1}{2} \alpha^2 Tr \lbrack \lambda_{(N+1)^2-1} \,, P \rbrack^2 \,,
\ee 
where $\lambda_{(N+1)^2-1}$ is the ``hypercharge'' generator of the global $U(N+1)$ symmetry of 
the NC ${\mathbb C}P^N$ model and $\alpha$ is a constant with dimensions of mass.
We observe that $V(P)$ breaks the global $U(N+1)$ symmetry of the pure NC ${\mathbb C}P^N$ model 
down to $U(1) \times U(N)$. The absolute minimum for the potential occurs at $P=0$, thus the global $U(1)$ 
symmetry is not spontaneously broken and $Q$-ball type solutions are possible \cite{Coleman, Leese}. Let us now see
how they come about.

The energy, topological charge and the Noether charge for the model may be given by the expressions:
\be
E = 2 \pi \theta Tr P \partial_i P \partial_i P + 2 \pi \theta \frac{1}{2}
Tr \partial_t P \partial_t P  - 2 \pi \theta \frac{1}{2}
\alpha^2 Tr \lbrack \lambda_{(N+1)^2-1} \,, P \rbrack^2 \,, 
\ee
\be
k = \frac{\theta}{i} \varepsilon_{ij} Tr P \partial_i P \partial_j P \,,
\ee
\be
Q = 2 \pi \theta i Tr \lambda_{(N+1)^2-1} \lbrack P \,, \partial_t P \rbrack \,.
\label{eq:Q}
\ee

Let us now consider the BPS-type inequality
\begin{multline}
2 \pi \theta \frac{1}{2} Tr (P \partial_\mu P \pm i \varepsilon_{\mu \nu} P \partial_\mu P)
(\partial_\mu P P \mp i \varepsilon_{\mu \rho} \partial_\rho P P ) \\
+ 2 \pi \theta \frac{1}{2} Tr (\partial_t P \pm i \alpha \lbrack \lambda_{(N+1)^2-1} \,, P \rbrack)
(\partial_t P \mp i \alpha \lbrack \lambda_{(N+1)^2-1} \,, P \rbrack^\dagger) \geq 0 \,.
\end{multline}
It implies immediately the bound
\be
E \geq 2 \pi |k| + \alpha |Q| \,,
\label{eq:energy}
\ee
which is saturated by the configurations satisfying
\be
P \partial_\mu P \pm i \varepsilon_{\mu \nu} P \partial_\mu P = 0 \,, \quad
\partial_t P \pm i \alpha \lbrack \lambda_{(N+1)^2-1} \,, P \rbrack = 0  \,.
\ee

We observe that the solutions of the above self-duality equations are given by the
BPS configurations of the NC ${\mathbb C}P^N$ model \cite{Bak}, which
now rotate with an angular frequency. The parameter $\alpha$ can be
rescaled for a given ${\mathbb C}P^N$ model with fixed $N$ such that the angular frequency
of rotations are $\alpha$. Then, the solutions can be specified by the partial isometry
\be
\chi(z \,, t) = 
\left (
\begin{array}{c}
u(z)e^{\pm i \alpha t} \\
1
\end{array}
\right ) 
\frac{1}{\sqrt{u^\dagger u + 1}} \,,
\label{eq:partialiso}
\ee 
where for self-dual solutions $u=u(z)$ is holomorphic in $z= x_1 +
x_2$ with $\lbrack z \,, {\bar z} \rbrack = 2 \theta$,
and anti-self-dual solutions are antiholomorphic $u=u({\bar z})$.

For the NC ${\mathbb C}P^1$ model we have
\be
P(u, u^\dagger, t) =
\left (
\begin{array}{cc}
u \frac{1}{u^\dagger u + 1} u^\dagger & u \frac{1}{u^\dagger u + 1} e^{\mp i \alpha t} \\
\frac{1}{u^\dagger u + 1} u^\dagger e^{\pm i \alpha t} & \frac{1}{u^\dagger u + 1} 
 \end{array}
\right ) \,,
\label{eq:qlump}
\ee
while in the ${\mathbb C}P^2$ model, $Q$-lumps are specified by the projector
\beqa
P(u, u^\dagger, t) = \left (
\begin{array}{ccc}
u_1 \frac{1}{\gamma} u_1^\dagger & u_1 \frac{1}{\gamma} u_2^\dagger & u_1 \frac{1}{\gamma} e^{\mp i \alpha t} \\
u_2 \frac{1}{\gamma} u_1^\dagger & u_2 \frac{1}{\gamma} u_2^\dagger & u_2 \frac{1}{\gamma} e^{\mp i \alpha t} \\
\frac{1}{\gamma} u_1^\dagger e^{\pm i \alpha t} & \frac{1}{\gamma} u_2^\dagger e^{\pm i \alpha t} & \frac{1}{\gamma}
\end{array} 
\right) 
\eeqa 
where $\gamma = u^\dagger_\alpha u_\alpha + 1$.

The potential terms in these models may also be expressed as 
\be
V(u) := \pi \theta \alpha^2 
Tr_{{\cal F}} \left ( u \frac{1}{(u^\dagger u + 1)^2} u^\dagger + \frac{u^\dagger u}{(u^\dagger u + 1)^2}  \right ) \,,
\label{eq:VCP1}
\ee
\be
V(u_1, u_2) = 2 \pi \theta \alpha^2 Tr_{{\cal F}} \left (u_1 \frac{1}{\gamma^2} u_1^\dagger +
u_2 \frac{1}{\gamma^2} u_2^\dagger + \frac{\gamma-1}{\gamma^2} \right ) \,.
\ee
These generalize the expression $V = \alpha^2 \int d^2 x  g^{\alpha \beta} u_\alpha u_\beta$, $g^{\alpha \beta}$ being the 
Fubini-Study metric on ${\mathbb C}P^N$, given in \cite{Abraham} for the ${\mathbb C}P^N$ models for the values $N=1,2$ and they collapse to it in the commutative limit.

It is apparent that the time dependence of these solutions is exactly the same as
that obtained in \cite{Leese, Abraham} for the $Q$-lumps of the
commutative ${\mathbb C}P^N$ models. Likewise, these solutions may have finite
energy only for winding numbers $k \geq 2$. This is due to the
fact that the Noether charge $Q$ diverges for configurations with $k = 1$, as
can be seen by inspecting the trace involved. We will return to the detailed
study of these traces shortly.

The contribution to the energy due the Noether charge lifts the degeneracy
of a class of solutions in the solution space of solitons of arbitrary sizes. 
This is quite expected as the addition of the potential term breaks the scaling 
invariance of the NC ${\mathbb C}P^N$ model even at the level of solutions\footnote{Recall that
the ${\mathbb C}P^N$ model action in NC spacetime is not scale invariant due to the noncommutativity,
however its solitonic solutions retain this feature.}. For instance, 
this is so for the radially symmetric configurations of the ${\mathbb C}P^1$ model : 
\be
u = \frac{(2 \theta)^{\frac{k}{2}}}{\lambda^k} a^k e^{i \alpha t } \,, \quad \lbrack a \,, a^\dagger \rbrack = 1 \,,
\quad k \neq 1 \,, \quad \lambda \neq 0 \,,
\label{eq:windingk}
\ee
where $\lambda$ characterizes the size of the soliton. These features are essentially the same as those found for the
commutative $Q$-lumps. 

However, it is important to remark that in contrast to the commutative theory, computing the Noether charge
or the energy for a given configuration with a generic winding number $k$ is not an easy task. Even for the class of winding 
number $k$ configurations specified by (\ref{eq:windingk}), it is rather difficult to compute the traces involved in $Q$. Explicitly, we have
\be
|Q| = 2 \pi \theta 2 \alpha \kappa^{2k} \Sigma_k(\kappa) = \pi \lambda^2 2 \alpha \kappa^{2k-2} \Sigma_k(\kappa) 
\,, \quad \kappa = \frac{\lambda}{\sqrt{2 \theta}} \,,
\label{eq:Q2nd}
\ee
where
\beqa
\Sigma_k (\kappa) &:=& 
\frac{1}{2} Tr_{\cal F} \left (\frac{a^{\dagger k} a^k}{\lbrack a^{\dagger k} a^k + \kappa^{2k} \rbrack^2} 
+ a^k \frac{1}{\lbrack a^{\dagger k} a^k + \kappa^{2k} \rbrack^2} a^{\dagger k}
\right) \nn \\
&=& \sum_{j=0}^\infty \frac{\frac{(j+k)!}{j!}}{\left(\frac{(j+k)!}{j!} + \kappa^{2k}  \right)^2} \,.
\label{eq:sigma}
\eeqa
The series $\Sigma_k(\kappa)$ converges for $ k \geq 2$ as can be verified by 
applying Raabe's test, while it diverges for $k=1$. 
For small values of $k$, the series $\Sigma_k (\kappa)$ may be summed by using Mathematica
or formulas from \cite{Gradshteyn}.
For $k=2$ we have\footnote{$\Sigma_3 (\kappa)$ is also available through Mathematica but significantly
more complicated then $\Sigma_2 (\kappa)$.},
\be
\Sigma_2 (\kappa) = \frac{\pi  \sec^2 \left(\frac{1}{2} \pi  \sqrt{1-4 \kappa ^4}\right) \left(-2 \pi  \kappa ^4 \sqrt{1-4 \kappa ^4}
+\left(1-2 \kappa ^4\right) \sin\left(\pi  \sqrt{1-4 \kappa ^4}\right)\right)}{2 \left(1-4 \kappa ^4\right)^{3/2}} \,.
\label{eq:sigma2}
\ee
This result will be made use of in the next section to concretely
demonstrate the new features encountered in the stability properties 
of NC $Q$-lumps under small radiative perturbations. 

The expression in (\ref{eq:Q2nd}) suggests that the Noether charge $Q$ is determined by $k, \alpha \,, \theta$ and $\lambda$.
In the commutative theory, $Q$ depends on $k, \alpha$ and $\lambda$ already \cite{Leese}, thus it is important to assess 
if and how the new alleged dependence of $Q$ on $\theta$ is genuine. For this purpose, let us first observe 
that for the solutions of the form (\ref{eq:windingk}) the moduli space metric is given by
\be
ds^2 = - \pi k^2 2^k \kappa^{2k-2}(t) \, \Sigma_k(\kappa(t)) \, (d \lambda^2) =: g_k(\kappa(t))  (d \lambda^2)  \,.
\label{eq:modulimetric}
\ee
Thus, it depends on $k$ and $\kappa(t)$ only. In order to claim that $Q$ indeed depends on $\theta$, 
there should be a way to fix our position in the moduli space while $\theta$ is still allowed to vary.
From (\ref{eq:modulimetric}) it is clear how this could be achieved. Namely, allowing $\theta$ and $\lambda$ to vary 
while keeping $\kappa$ fixed, the moduli space metric does not change; however, $Q$ continues to
vary with $\theta$. In other words, taking $\theta$ proportional to $\lambda^2$, we can think of $Q$ as a function of
either $\theta$ or $\lambda$. It is only in this regime that $Q$ (and consequently the energy) may be viewed to depend 
on $\theta$. 

It is also important to study the case when $\theta$ is fixed and 
$\lambda$ is allowed to vary. In this situation, although both the metric $g_k(\kappa)$ and $Q$ continue to 
vary with $\lambda$, we notice from (\ref{eq:Q2nd}) and (\ref{eq:modulimetric}) that we can always write
$Q = - \frac{\alpha}{2^{k-1} k^2} \lambda^2 g_k(\kappa) $, and thus the factor $g_k(\kappa)$ entirely compensates for the change 
in the moduli space metric as $\lambda$ is varied and consequently we can view $Q$ as a function of $\lambda$. 

Finally, we note that the commutative limit is recovered by taking $\kappa \rightarrow \infty$, keeping $\lambda$ fixed 
and taking $\theta \rightarrow 0$.

\section{Stability of NC ${\mathbb C}P^1$ $Q$-lumps}

Elementary classical stability properties of the NC $Q$-lumps are also
quite similar to their commutative counterparts. As the NC $Q$-lumps
saturate the BPS-type bound, they are automatically classically
stable configurations. The quantum stability of $Q$-lumps requires
the energy-charge ratio to be smaller than the meson mass in the
theory \cite{Leese}. For the present case, the maximum energy transferable to
radiating mesons is $E^\prime = E - 2 \pi N$ due to the topological
stability of the NC $Q$-lump configuration, and hence the
energy-charge ratio is given as $\frac{E^\prime}{Q} = \alpha$, 
due to the BPS-type bound. This is precisely the same situation
encountered for the commutative $Q$-lumps, which appear to be at the 
threshold of quantum stability. Consequently, the same
crucial question regarding the radiative stability of the $Q$-lumps, 
which was analyzed in detail by Leese \cite{Leese}, is also of
essential interest here. To be more precise, although the NC $Q$-lumps 
appear to be quantum mechanically stable (more accurately at the
threshold of quantum stability), it is easily seen that any small 
perturbation could start a continuous emission of radiating
mesons as the ratio $\frac{E^\prime}{Q}$ grows larger once such a
process is initiated, and this would eventually lead the $Q$-lump to 
shrink to a spike. Therefore, the question which needs to be answered is
whether such a continuous emission of radiating mesons is classically
possible. Leese analyzed this problem using numerical techniques and also provided a rather
simple analytic discussion that corroborates with his numerical findings that such radiative
instabilities are not present as long as there is a potential barrier between $Q$-lumps with winding number
$k$ and configurations consisting of $Q$-lumps with winding number $k^\prime < k$ present together with
some mesons at larger distances. The situation in the noncommutative
setting appears to be somewhat more complicated, and at present we
will not attempt to give a full result using numerical techniques. In what follows,
we apply the aforementioned analytical procedure to the NC
$Q$-lumps. This will help us to see some new features of these
configurations and also allow us to show that they remain stable
under small radiative perturbations.

The most general radially symmetric configuration (not necessarily a solution) with winding number $k$ may be given by
\be
u = \frac{(2\theta)^{\frac{k}{2}}}{\lambda^k({\hat N}\,, t)} a^k e^{i \psi({\hat N} \,, t)} \,, 
\quad \lambda({\hat N}\,, t) \neq 0 \,.
\label{eq:radial}
\ee
In (\ref{eq:radial}), ${\hat N} = a^\dagger a$ is the number operator, which maps to the square of the radial
coordinate under the diagonal coherent states map. Assuming that the system remains approximately radially symmetric
during the time evolution, we can drop the ${\hat N}$ dependence in
$\lambda({\hat N} \,, t)$ and $\psi({\hat N} \,, t)$. Then, the associated Lagrangian becomes
\be
L_k = 2 \pi k + 2 \pi \theta \frac{\lambda^{2k}(t)}{\theta^k} \left (
\alpha^2 - \left
( \frac{\lambda(t)^{\prime 2}}{\lambda(t)^2} k^2 + \psi^{\prime 2}(t) \right ) \right ) 
\Sigma_k \left(\frac{\lambda(t)}{\sqrt{2 \theta}} \right) \,,
\label{eq:Lagrangiank}
\ee
where $^\prime$ denotes the derivatives with respect to $t$.

$L_k(\lambda(t) \,, \psi(t))$ specifies a dynamical system in $\lambda(t)$ and $\psi(t)$. Let us now consider 
a small perturbation around the $Q$-lump configuration (\ref{eq:qlump})
\be 
\lambda(t) = \lambda_0  + \varepsilon(t) \,, \quad \psi(t) = t ( \alpha + \delta(t)) \,,
\label{eq:perturbation}
\ee
where $\lambda_0$ stands for $\lambda(t=0)$ for short. Using the equation of motion for $\psi(t)$ and 
(\ref{eq:perturbation}) we have
\be
\psi^\prime(t) = \alpha \frac{\lambda_0^{2k} 
\Sigma_k \left(\frac{\lambda_0}{\sqrt{2
      \theta}}\right)}{\lambda^{2k}(t) 
\Sigma_k \left(\frac{\lambda(t)}{\sqrt{2 \theta}}\right)} \,.
\ee
For our purposes, we only need the equation of motion\footnote{The full 
equation of motion is given in the appendix for completeness.}
for $\lambda(t)$ at first order in $\varepsilon(t)$.
This takes the form
\be
\varepsilon^{\prime \prime}(t) + \frac{4 \alpha^2}{k^2} A_k^2(\kappa_0) \varepsilon(t) + O (\varepsilon^2(t))  = 0 \,, 
\ee
where 
\be
A_k(\kappa) := \left( k + \frac{1}{2} \kappa_0 \frac{\partial_\kappa \Sigma_k(\kappa) \big|_{\kappa = \kappa_0}}
{\Sigma_k(\kappa_0)} \right ) \,, \quad \kappa_0 = \frac{\lambda_0}{\sqrt{2 \theta}} \,.
\ee
Thus, the width of the radially symmetric lumps oscillate under small 
perturbations albeit with a period $\tau = \frac{\pi k}{\alpha
A_k(\kappa_0)}$, which now depends on the
function $A_k(\kappa_0)$ of $\kappa_0 = \frac{\lambda_0}{\sqrt{2
\theta}}$ in contrast to the commutative theory. We conclude that the $Q$-lump configurations remain stable under small
radiative perturbations. As $\kappa_0$ gets larger (i.e. $\theta \ll \lambda_0$), 
we have that $\Sigma_k(\kappa_0 \rightarrow \infty) \approx \kappa_0^{2-2k}$ and hence
$A_k(\kappa_0 \rightarrow \infty) \rightarrow 1$. 
The commutative limit is, therefore, smoothly recovered.

It is possible to investigate the behavior of $A_{k=2}(\kappa_0)$ in
more detail. In Fig. 1, we have plotted the response of $A_2(\kappa_0)$.
We observe that $A_2(\kappa_0)$ smoothly approaches to the value $1$  as $\kappa_0 \rightarrow \infty$ 
and hence $\tau$ approaches its commutative value. Put another way, we could state that 
the noncommutativity leads to more rapid oscillations around the $Q$-lump
configuration. 
\begin{figure}  
\begin{center}
\includegraphics[width=0.60\textwidth, height=0.25\textheight]{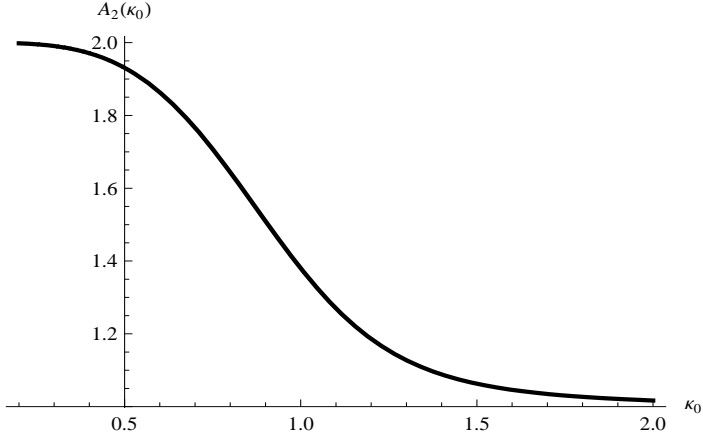}
\caption{Plot of $A_2(\kappa_0)$ as a function of $\kappa_0$.}
\end{center}
\end{figure}

\section{Conclusions and Outlook}

In this paper we have constructed the $Q$-lump configurations associated with the NC 
${\mathbb C}P^N$ models. We have found that, similar to their commutative counterparts, they
too appear as extended field configurations, which rotate with a fixed angular frequency in 
time, and saturate a BPS-type bound. Quite interestingly, it was also found that in a regime in which 
$\theta$ is taken to be proportional to $\lambda^2$, the energy of the NC $Q$-lump configurations 
can be viewed to depend on $\theta$ via the Noether charge $Q$. A collective coordinate-type analysis helped us to show that ${\mathbb C}P^1$ model $Q$-lumps
remain stable under small radiative perturbations. Contrary to the behavior of the commutative $Q$-lumps, we 
have also seen that, due to the noncommutativity the period of fluctuations around the radially symmetric $Q$-lump 
configurations depend on the function $A(\kappa_0)$ leading to more rapid oscillations around the $Q$-lump
configuration. 

It seems rather straightforward to obtain the supersymmetric extensions of the NC $Q$-lumps following the ideas of 
\cite{Alvarez-Gaume-Freedman, Bak}. 
To be more concrete, focusing on the ${\cal N} =2$ superspace ${\mathcal A}_\theta({\mathbb R}^{2+1 \, | 4})$ with only the Moyal-type noncommutativity, (i.e. Grassmann coordinates are undeformed and they anticommute), it is possible to consider the supersymmetric Lagrangian
\be
L = \int d^2 \theta \, Tr D {\cal P} {\bar D} {\cal P}  + {\cal W}(\chi) \,,
\ee
where ${\cal P} \equiv {\cal P}({\hat x}_\mu, \theta_\alpha)$ is a projector in ${\mathcal A}_\theta({\mathbb R}^{2+1 \, | 4})
\otimes Mat(N+1)$, $\chi \equiv \chi({\hat x}_\mu, \theta_\alpha)$ is the partial isometry fulfilling 
${\cal P} = \chi \chi^\dagger$ and $\chi^\dagger \chi = 1$ and ${\cal W}(\chi)$ is a superpotential. Following \cite{Bak}, the superpotential can be taken to be of the form
\be
{\cal W}(\chi) = \beta Tr \chi^\dagger {\cal K} \chi \,,
\ee
where ${\cal K} = diag(1,1,1, \cdots, 0) \in Mat(N+1)$ is a projector. It may be shown that this system leads to the same bosonic NC ${\mathbb C}P^N$ $Q$-lump configurations with half of the supersymmetries, while the fermionic part is assumed to vanish. 
Clearly, a comprehensive study of noncommutative deformations of massive supersymmetric sigma models in various dimensions still 
has to be made to shed more light into their detailed structure.

There are several other issues which remain to be investigated. First of all, it should be possible to 
explore the scattering of NC $Q$-lumps and compare it with those of the commutative theory, 
as well as with those in the pure NC ${\mathbb C}P^N$ models. It may also be possible to explore the addition of topological terms, 
such as the Chern-Simons term or a Berry phase like term into the action. It appears that the latter of these lead to 
divergent contributions in general \cite{unpublished}. Nevertheless, it may be possible to regulate the contribution of divergent
traces by restricting the configuration space to an infinite strip \cite{Seckin} or a disc \cite{Lizzi} on the GM plane. Such a regularization, however, also alters the structure of the pure sigma model lumps as well as the $Q$-lumps, and further investigation is necessary to understand the behavior of these configurations. We hope to report on the progress on these topics elsewhere. 

\vskip 2em 

{\bf Acknowledgments} 

\vskip 1em

I thank O. Lechtenfeld for a careful reading of the manuscript and critical comments and suggestions.
This work is supported by the Deutsche Forschungsgemeinschaft (DFG) under Grant No. LE 838/9. 

\appendix

\section{Appendix}

The equation for $\lambda(t)$ is
\be
\lambda^{\prime \prime}(t) + \frac{1}{2} \Big( 2(k-1) + \kappa(t) \frac{\partial_\kappa \Sigma_k(\kappa) }
{\Sigma_k(\kappa)} \Big) 
\frac{\lambda^{\prime 2}(t)}{\lambda(t)} + \frac{1}{k^2} \Big(\alpha^2 - \psi^{\prime 2}(t) \Big) \left ( k + \frac{1}{2} \kappa(t)
\frac{\partial_\kappa \Sigma_k(\kappa) }
{\Sigma_k(\kappa)} \right ) \lambda(t) = 0 \,.
\ee
As $\theta \rightarrow 0$, $\kappa(t)\frac{\partial_\kappa \Sigma_k(\kappa)} {\Sigma_k(\kappa)} \rightarrow 2(1-k)$ and 
the result of the commutative theory is recovered.


\begin{thebibliography}{99}

\bibitem{Leese} R.~A.~Leese,
  ``Q lumps and their interactions,''
  Nucl.\ Phys.\  B {\bf 366}, 283 (1991).

\bibitem{Coleman} S.~R.~Coleman,
  ``Q Balls,''
  Nucl.\ Phys.\  B {\bf 262}, 263 (1985)
  [Erratum-ibid.\  B {\bf 269}, 744 (1986)].

\bibitem{Abraham} E.~Abraham,
  ``Nonlinear Sigma Models And Their Q Lump Solutions,''
  Phys.\ Lett.\  B {\bf 278}, 291 (1992).

\bibitem{Abraham-Townsend} E.~R.~C.~Abraham and P.~K.~Townsend,
  ``Q kinks,''
  Phys.\ Lett.\  B {\bf 291}, 85 (1992);

E.~R.~C.~Abraham and P.~K.~Townsend,
  ``More On Q Kinks: A (1+1)-Dimensional Analog Of Dyons,''
  Phys.\ Lett.\  B {\bf 295}, 225 (1992).

\bibitem{Gauntlett-Tong-Townsend}J.~P.~Gauntlett, D.~Tong and P.~K.~Townsend,
  ``Multi-domain walls in massive supersymmetric sigma-models,''
  Phys.\ Rev.\  D {\bf 64}, 025010 (2001)
  [arXiv:hep-th/0012178];

J.~P.~Gauntlett, R.~Portugues, D.~Tong and P.~K.~Townsend,
  ``D-brane solitons in supersymmetric sigma-models,''
  Phys.\ Rev.\  D {\bf 63}, 085002 (2001)
  [arXiv:hep-th/0008221].

\bibitem{Naganuma-Nitta-Sakai}M.~Naganuma, M.~Nitta and N.~Sakai,
  ``BPS lumps and their intersections in N = 2 SUSY nonlinear sigma models,''
  Grav.\ Cosmol.\  {\bf 8}, 129 (2002)
  [arXiv:hep-th/0108133];

M.~Eto, Y.~Isozumi, M.~Nitta and K.~Ohashi,
  ``1/2, 1/4 and 1/8 BPS equations in SUSY Yang-Mills-Higgs systems: Field
  theoretical brane configurations,''
  Nucl.\ Phys.\  B {\bf 752}, 140 (2006)
  [arXiv:hep-th/0506257].

\bibitem{Bak} D.~Bak, S.~O.~Hahn, J.~Lee and P.~Oh,
  ``Supersymmetric Q-lumps in the Grassmannian nonlinear sigma models,''
  Phys.\ Rev.\  D {\bf 75}, 025004 (2007)
  [arXiv:hep-th/0610067].

\bibitem{Nekrasov}  M.~R.~Douglas and N.~A.~Nekrasov,
  ``Noncommutative field theory,''
  Rev.\ Mod.\ Phys.\  {\bf 73}, 977 (2001) [arXiv:hep-th/0106048].

\bibitem{Szabo} R.~J.~Szabo,
  ``Quantum field theory on noncommutative spaces,''
  Phys.\ Rept.\  {\bf 378}, 207 (2003)
  [arXiv:hep-th/0109162].

\bibitem{Harvey1}
  J.~A.~Harvey,
  ``Komaba lectures on noncommutative solitons and D-branes,''
  arXiv:hep-th/0102076.

\bibitem{Fuzzy} A.P. Balachandran, S. K\"{u}rk\c{c}\"{u}o\v{g}lu, S. Vaidya, 
{\it Lectures on Fuzzy and Fuzzy SUSY Physics}, World Scientific, Singapore, 2007, and hep-th/0511114.

\bibitem{Lee} B.~H.~Lee, K.~M.~Lee and H.~S.~Yang,
  ``The CP(n) model on noncommutative plane,''
  Phys.\ Lett.\  B {\bf 498}, 277 (2001)
  [arXiv:hep-th/0007140].

\bibitem{Lechtenfeld-Domrin} A.~V.~Domrin, O.~Lechtenfeld and S.~Petersen,
  ``Sigma-model solitons in the noncommutative plane: Construction and
  stability analysis,''
  JHEP {\bf 0503}, 045 (2005)
  [arXiv:hep-th/0412001].

\bibitem{QBalls} Y.~Kiem, C.~j.~Kim and Y.~b.~Kim,
  ``Noncommutative Q-balls,''
  Phys.\ Lett.\  B {\bf 507}, 207 (2001)
  [arXiv:hep-th/0102160].

\bibitem{Gradshteyn} I.S. Gradshteyn and I.M. Ryzhik, {\it Table of Integrals Series and Products}, 
Academic Press, London, 1965.

\bibitem{Alvarez-Gaume-Freedman} L.~Alvarez-Gaume and D.~Z.~Freedman,
  ``Potentials For The Supersymmetric Nonlinear Sigma Model,''
  Commun.\ Math.\ Phys.\  {\bf 91}, 87 (1983).

\bibitem{unpublished} A.P. Balachandran, K.S. Gupta, S.K\"{u}rk\c{c}\"{u}o\v{g}lu, Unpublished.

\bibitem{Seckin} A.~P.~Balachandran, K.~S.~Gupta and S.~Kurkcuoglu,
  ``Edge currents in non-commutative Chern-Simons theory from a new matrix
  model,''
  JHEP {\bf 0309}, 007 (2003)
  [arXiv:hep-th/0306255].

\bibitem{Lizzi} F.~Lizzi, P.~Vitale and A.~Zampini,
  ``The fuzzy disc,''
  JHEP {\bf 0308}, 057 (2003)
  [arXiv:hep-th/0306247]; 

F.~Lizzi, P.~Vitale and A.~Zampini,
  ``The beat of a fuzzy drum: Fuzzy Bessel functions for the disc,''
  JHEP {\bf 0509}, 080 (2005)
  [arXiv:hep-th/0506008].

\end{thebibliography}
\end{document}